\documentclass[11pt,a4paper]{article}

\usepackage[truedimen,margin=34mm]{geometry} 

\usepackage{mathrsfs}
\usepackage{amssymb}
\usepackage{amsmath}
\usepackage{ascmac}
\usepackage{amsthm}
\usepackage[pdftex]{graphicx}
\usepackage{booktabs}
\usepackage{setspace}
\usepackage{natbib}
\usepackage{color}

\usepackage{comment}

\usepackage{titlesec}
\titleformat*{\section}{\large\bfseries}
\titleformat*{\subsection}{\it}

\def\ep{{\varepsilon}}
\def\la{{\lambda}}

\title{{\bf Robust Bayesian Regression with \\ Synthetic Posterior}}

\date{}
\author{}

\begin{document}

\maketitle
\doublespacing

\vspace{-1.5cm}
\begin{center}
{\large Shintaro Hashimoto$^1$ and Shonosuke Sugasawa$^2$}
\end{center}

\medskip
\noindent
$^1$Department of Mathematics, Graduate School of Science, Hiroshima University\\
$^2$Center for Spatial Information Science, The University of Tokyo

\vspace{5mm}
\begin{center}
{\bf \large Abstract}
\end{center}
Although linear regression models are fundamental tools in statistical science, the estimation results can be sensitive to outliers.
While several robust methods have been proposed in frequentist frameworks, statistical inference is not necessarily straightforward. 
We here propose a Bayesian approach to robust inference on linear regression models using synthetic posterior distributions based on $\gamma$-divergence, which enables us to naturally assess the uncertainty of the estimation through the posterior distribution. 
We also consider the use of shrinkage priors for the regression coefficients to carry out robust Bayesian variable selection and estimation simultaneously. 
We develop an efficient posterior computation algorithm by adopting the Bayesian bootstrap within Gibbs sampling.
The performance of the proposed method is illustrated through simulation studies and applications to famous datasets. 


\newpage

\section{Introduction}
\label{}

Linear regression models are fundamental tools in statistical science. 
However, it is recognized that the model can be highly influenced by outliers, which may result in biased or inefficient statistical inference on regression coefficients and an error variance. 
In the content of frequentist inference, general robust estimation methods using divergence \citep[e.g.][]{B1998, FE2008, Jones2001} are known to be appealing, and specialized methods for regression models have been proposed \citep{KF2017, KF2019}. 
On the other hand, the valid inference under existing outliers would be a challenging problem even if the divergence method is adopted, and the problem is more difficult when the penalized methods for variable selection such as Lasso \citep{Tib1996} are incorporated.

In this paper, we develop a Bayesian approach for robust inference in linear regression models based on $\gamma$-divergence \citep{Jones2001, FE2008}.
We adopt a framework of synthetic (general) posterior inference \citep[e.g.][]{Bissiri2016, Jewson2018, BPY2019, MD2019, Nakagawa2019}, and define the synthetic posterior distribution of the unknown parameters in the linear regression models by replacing the log-likelihood function with $\gamma$-divergence, which enables us to naturally carry out point estimation as well as uncertainty quantification based on the posterior. 
For the prior distributions of the regression coefficients, we assign a class of shrinkage priors expressed as a scale mixture of normals that includes typical ones such as Laplace \citep{PC2008} and horseshoe \citep{Cal2010} priors, leading to robust Bayesian variable selection and estimation at the same time. 
However, the main difficulty of using the proposed synthetic posterior is that the form of the posterior is a complicated function of the unknown parameters, thereby the efficient posterior computation algorithms as done in the standard Bayesian linear regression models are no more applicable. 
To solve the issue, we develop an efficient sampling algorithm using the Bayesian bootstrap \citep[e.g.][]{R1981, NR1994, Lyddon2018, NPX18} within Gibbs sampling. 
In the step of the Bayesian bootstrap, we maximize the randomized posterior distribution to generate a sample from the full conditional distributions of the unknown parameters, which will be shown to be efficiently carried out by modifying the Majorization-Minimization (MM) algorithm given in \cite{KF2017}.
The advantage of the proposed algorithm is that there is no rejection steps that are typically required in the standard Metropolis-Hastings algorithm, and performance in terms of mixing and autocorrelation are quite reasonable as shown in our numerical results.

In the context of Bayesian inference, standard approaches for robust inference are replacing the normal distribution with heavy-tailed distributions for the error term. 
The most typical choice is the $t$-distribution, but this approach is not necessarily a good solution and still suffer from undesirable performance under some scenarios of outliers as shown in our simulation studies. 
Recently, \cite{Gag2020} proposed a more robust error distribution than the Cauchy distribution, but the new distribution does not admit stochastic representations that facilitate efficient posterior computation, which would be quite problematic especially when the number of covariates is large as in most modern applications.  
Moreover, the advantage of the synthetic posterior approach compared with the use of heavy-tailed distribution is that the synthetic posterior can still be applicable to other types of regression models such as logistic and Poisson regressions although we focus only on the application to linear regression in this paper.

The paper is organized as follows: In Section \ref{sec:meth}, we introduce the setting of our model and formulate the synthetic posterior distribution. Next, we consider the Bayesian linear regression model under shrinkage priors expressed as scale mixtures of normals. Then a new posterior computation algorithm for synthetic posterior with shrinkage priors is proposed.  
In Section \ref{sec:num}, we provide results of numerical studies and real data analysis. Under several types of contaminations, it is shown that the proposed method outperforms the original Bayesian lasso and Bayesian lasso with the $t$-distribution as the error distribution. Also, we show mixing properties of the proposed algorithm compared with other options.


\section{Robust Bayesian regression via synthetic posterior}
\label{sec:meth}

\subsection{Settings and synthetic posterior}\label{sec:model}
Suppose we have independent observation $(y_i,x_i)$ for $i=1,\ldots,n$, where $y_i$ is a continuous response and  $x_i=(x_{i1},\ldots,x_{ip})^{\top}$ is a $p$-dimensional vector of covariates.
We consider fitting a regression model $y_i=x_i^{\top}\beta+\ep_i$ with $\ep_i\sim N(0,\sigma^2)$.
Let $\pi(\beta,\sigma^2)$ be a prior distribution for the model parameters $\beta$ and $\sigma^2$.
Then the standard posterior distribution of $(\beta, \sigma^2)$ is given by 
\begin{equation}\label{pos}
\pi(\beta,\sigma^2 | D)=\pi(\beta,\sigma^2)\exp\left\{\sum_{i=1}^n\log f(y_i;x_i^{\top} \beta,\sigma^2)\right\},
\end{equation}
where $f(y_i;x_i^{\top} \beta,\sigma^2)$ is the density function of $N(x_i^{\top} \beta,\sigma^2)$, and $D$ denotes the set of sampled data.
When there exist outliers which have large residuals $(y_i-x_i^{\top} \beta)/\sigma$, the posterior distribution (\ref{pos}) is known to be sensitive to such outliers, and it can produce biased or inefficient posterior inference.

To overcome the problem, we propose replacing the log-likelihood function in (\ref{pos}) with robust alternatives.
Specifically, we employ $\gamma$-divergence \citep{Jones2001, FE2008} of the form:
\begin{equation*}
R_{\gamma}(\beta,\sigma^2)=\frac{n}{\gamma}\log\left\{\frac1n\sum_{i=1}^n\left(\frac{f(y_i;x_i^{\top} \beta,\sigma^2)}{\|f(\cdot;x_i^{\top} \beta,\sigma^2)\|_{\gamma+1}}\right)^{\gamma}\right\},
\end{equation*}
where
$$
\|f(\cdot;x_i^{\top} \beta,\sigma^2)\|_{\gamma+1}=\left(\int f(t;x_i^{\top} \beta,\sigma^2)^{1+\gamma}\mathrm{d}t\right)^{1/(1+\gamma)}.
$$
Note that $\gamma$ is a tuning parameter that controls the robustness, and $R_{\gamma}(\theta)$ reduces to the log-likelihood function as $\gamma\to 0$. We now define the following synthetic posterior based on $\gamma$-divergence:
\begin{equation}\label{SP}
\begin{split}
&\pi_{\gamma}(\beta,\sigma^2 | D)
\propto \pi(\beta,\sigma^2)\exp\left\{R_{\gamma}(\beta,\sigma^2)\right\}.
\end{split}
\end{equation} 
Since $\pi_{\gamma}(\beta,\sigma^2 | D)$ reduces to the standard posterior (\ref{pos}) under $\gamma\to 0$, the synthetic posterior (\ref{SP}) can be regarded as a natural extension of the standard posterior (\ref{pos}).

A robustness property of the posterior distribution (\ref{SP}) can be checked by considering the framework adopted in \cite{Gag2020}. 
Let $\ell_i$ be an indicator of being non-outlying observations, that is, $\ell_i=1$ if $y_i$ is not an outlier and $\ell_i=0$ if $y_i$ is an outlier.
For outliers, we consider a situation that $y_i=a_i+b_i\omega$ and $\omega\to\infty$. 
Under the framework, it follows that $f(y_i;x_i^{\top} \beta,\sigma^2)^{\gamma}\to 0$ for arbitral $(\beta,\sigma^2)\in \Theta$, where $\Theta$ is a compact set.
Therefore, the posterior distribution (\ref{SP}) converges to the distribution proportional to 
$$
\pi(\beta,\sigma^2)\exp\left[\frac{n}{\gamma}\log\left\{\frac1{n_\ell}\sum_{i: \ell_i=1}\left(\frac{f(y_i;x_i^{\top} \beta,\sigma^2)}{\|f(\cdot;x_i^{\top} \beta,\sigma^2)\|_{\gamma+1}}\right)^{\gamma}\right\}\right],  \ \ \ (\beta,\sigma^2)\in \Theta,
$$
as $\omega\to\infty$, where $n_\ell=\sum_{i=1}^n\ell_i$ is the number of non-outlying observations. 
This means that the information of outliers are automatically ignored in the posterior distribution (\ref{SP}) as long as the outlier values are extreme, that is, the outliers and non-outliers are well-separated.

\subsection{Posterior computation}
We first consider the standard prior for $(\beta,\sigma^2)$, namely, normal prior for $\beta$ and inverse gamma prior for $\sigma^2$, independently, given by $\pi(\beta,\sigma^2)\propto \exp(-\beta^{\top} S_\beta^{-1}\beta/2)(\sigma^2)^{-a/2-1}\exp\{-a/(2\sigma^2)\}$,
where $S_\beta$ and $a$ are hyperparameters. 
Since the synthetic posterior distribution (\ref{SP}) is not a familiar form, the posterior computation to generate random samples of $(\beta,\sigma^2)$ is not straightforward.
We use crude approaches such as Metropolis-Hastings algorithm, but the acceptance probabilities might be very small even when the dimension of covariates $x_i$ is moderate.
To avoid such undesirable situation, we adopt approximated sampling strategy using weighted likelihood bootstrap \citep{NR1994} which generate posterior samples as the minimizer of the weighted objective function:
\begin{equation}\label{NB}
\begin{split}
L_w(\beta,\sigma^2)
&=-\frac{n}{\gamma}\log\left\{\frac1n\sum_{i=1}^nw_if(y_i; x_i^{\top} \beta,\sigma^2)^{\gamma}\right\}+\frac12\beta^{\top} S_\beta^{-1}\beta\\
& \ \ \ +\left(1+\frac{a}{2}-\frac{n\gamma}{2(1+\gamma)}\right)\log\sigma^2+\frac{a}{\sigma^2},
\end{split}
\end{equation}
where $(w_1,\ldots,w_n)\sim n\cdot {\rm Dirichlet}(1,\ldots,1)$, so that $\sum_{i=1}^nw_i=n$.
The minimization of (\ref{NB}) can be efficiently carried by MM algorithm \citep{HL2004} obtained by slight modification of one given in \cite{KF2017}. 
From Jensen's inequality, with current values $(\beta_{\ast},\sigma_{\ast}^2)$ of the model parameters, the upper bound of the objective function (\ref{NB}) can be obtained as follows: 
\begin{equation}\label{NB-upper}
\begin{split}
L_w(\beta,\sigma^2)
&=\frac1{2\sigma^2}\sum_{i=1}^ns_i^{\ast} (y_i-x_i^{\top} \beta)^2+\frac12\beta^{\top} S_\beta^{-1}\beta\\
&\ \ \ +\left(1+\frac{a}{2}+\frac{n}{2(1+\gamma)}\right)\log\sigma^2+\frac{a}{\sigma^2}+C,
\end{split}
\end{equation}
where $C$ is an irrelevant constant and $s_i^{\ast}$ is a new weight defined as  
\begin{equation}\label{weight}
s_i^{\ast}=\frac{w_if(y_i; x_i^{\top} \beta_{\ast},\sigma_{\ast}^2)^{\gamma}}{\sum_{j=1}^nw_jf(y_j; x_j^{\top} \beta_{\ast},\sigma_{\ast}^2)^{\gamma}},
\end{equation}
noting that $\sum_{i=1}^n s_i^{\ast}=n$.
The upper bound in (\ref{NB-upper}) can be easily minimized, so that the updating process is given by 
\begin{equation}\label{update}
\begin{split}
&\beta_{\dagger}=\left\{\frac1{\sigma_{\ast}^2}\sum_{i=1}^ns_i^{\ast}x_ix_i^{\top} +S_\beta^{-1}\right\}^{-1}\frac1{\sigma_{\ast}^2}\sum_{i=1}^ns_i^{\ast}x_iy_i,\\
&\sigma^2_{\dagger}=\left(2+a+\frac{n}{1+\gamma}\right)^{-1}\left\{a+\sum_{i=1}^ns_i^{\ast}(y_i-x_i^{\top} \beta_{\dagger})^2\right\}.
\end{split}
\end{equation}
Therefore, the MM algorithm to get the minimizer of (\ref{NB}) repeats calculation of the weight (\ref{weight}) and updating parameter values via (\ref{update}) until convergence.

For generating $B$ random samples from synthetic posterior distribution (\ref{SP}), we generate $B$ samples of $w_i$'s and solve minimization problems of (\ref{NB}) for $B$ times via the MM algorithm.
The resulting samples can be approximately regarded as posterior samples of (\ref{SP}), and the approximation error would be negligible when $n$ is large \citep[e.g.][]{Lyddon2018, NPX18}.

\subsection{Incorporating shrinkage priors}
When the dimension of $x_i$ is moderate or large, it is desirable to select a subset of $x_i$ that are associated with $y_i$, which corresponds to shrinkage estimation of $\beta$.
Here we rewrite the regression model to explicitly express an intercept term as $y_i=\alpha+x_i^{\top}\beta+\ep_i$.
We consider normal prior for $\alpha$, that is, $\alpha\sim N(0,S_{\alpha})$ with fixed $S_{\alpha}>0$.
For coefficients $\beta$, we introduce shrinkage priors expressed as a scale mixture of normals given by 
\begin{equation}\label{prior}
\pi(\beta)=\prod_{k=1}^p\int_{0}^\infty \phi(\beta_k; 0,u_k)g(u_k;\la){\rm d}u_k,
\end{equation}
where $g(\cdot;\la)$ is a mixing distribution which may depends on some tuning (scale) parameter $\la$. 
Many existing shrinkage priors are included in this class by appropriately choosing the mixing distribution $g(\cdot)$.
Among many others, we consider two priors for $u_i$: exponential distribution which results in Laplace prior of $\beta$ known as Bayesian Lasso \citep{PC2008}, and half-Cauchy distribution for $\sqrt{u_k}$ which results in horseshoe prior for $\beta$ \citep{Cal2010}.
The detailed settings of these two priors are given as follows:
\begin{align*}
{\rm (Laplace)}& \ \ \ u_k|\la\sim {\rm Exp}(\la^2/2), \ \ \ \la^2\sim {\rm Ga}(c_1,c_2)\\
{\rm (Horseshoe)}& \ \ \ u_k|\xi_k,\la\sim {\rm IG}(1/2,\la/\xi_k), \ \ \ \xi_k\sim {\rm IG}(1/2,1), \ \ \ \la\sim {\rm Ga}(c_1,c_2),
\end{align*}
where $c_1$ and $c_2$ are fixed hyperparameters, ${\rm Ga}(a,b)$ denotes the gamma distribution with shape parameter $a$ and rate parameter $b$, and ${\rm IG}(a,b)$ is the inverse gamma distribution with shape parameter $a$ and scale parameter $b$.
Note that $\xi_k$ in the horseshoe prior is additional latent variable to make posterior computation easier, and the density of $u_k|\la$ is proportional to $u_k^{-1/2}(u_k+\la)^{-1}$, so that $\sqrt{u_k}|\la$ follows half-Cauchy distribution. 

Using the mixture representation (\ref{prior}), the full conditional distribution of $(\alpha,\beta,\sigma^2)$ given the other parameters including $u_k$'s is given by
\begin{equation*}
\begin{split}
&\exp\left(-\frac{\alpha^2}{2S_{\alpha}}-\frac12\beta^{\top}U^{-1}\beta\right)(\sigma^2)^{-a/2-1}\exp\left(-\frac{a}{2\sigma^2}\right)\\
& \ \ \ \times \exp\left[\frac{n}{\gamma}\log\left\{\frac1n\sum_{i=1}^nf(y_i; \alpha+x_i^{\top}\beta,\sigma^2)^{\gamma}\right\}+\frac{n\gamma}{2(1+\gamma)}\log\sigma^2\right],
\end{split}
\end{equation*}
where $U={\rm diag}(u_1,\ldots,u_p)$.
Similarly to the previous section, we can generate approximate posterior samples of $(\alpha,\beta,\sigma^2)$ by minimizing the weighted objective function obtained by replacing $f(y_i; \alpha+x_i^{\top}\beta,\sigma^2)^{\gamma}$ with $w_if(y_i; \alpha+x_i^{\top}\beta,\sigma^2)^{\gamma}$ in the above expression, where $w_i$ is defined in the same way in the previous section.
Then, the objective function can be minimized by a similar MM-algorithm given in the previous section.
Under given regression coefficients $\beta$, the full conditional distribution of latent variables $u_1,\ldots,u_p$ and hyperparameter $\la$ are the same as the case with the standard linear regression, so that we can use the existing Gibbs sampling methods.
We summarize our Markov chain Monte Carlo (MCMC) algorithm under the two shrinkage priors in what follows.

\vspace{5mm}
\noindent
{\bf MCMC algorithm under shrinkage prior}
\begin{itemize}
\item
(Sampling from $\alpha,\beta$ and $\sigma^2$) \ \ \ Generate $w_i=n\cdot {\rm Dir}(1,\ldots,1)$, and set initial values $\alpha_{(0)},\beta_{(0)}$ and $\sigma^2_{(0)}$.
Repeat the following procedures until convergence:
\begin{itemize}
\item
Compute the following weight:
\begin{equation*}
s_i^{(k)}=\frac{w_if(y_i; \alpha_{(k)}+x_i^{\top}\beta_{(k)},\sigma_{(k)}^2)^{\gamma}}{\sum_{j=1}^nw_jf(y_j; \alpha_{(k)}+x_j^{\top}\beta_{(k)},\sigma_{(k)}^2)^{\gamma}}, \ \ \ i=1,\ldots,n.
\end{equation*}

\item
Update the parameter values:
\begin{equation*}
\begin{split}
&\alpha_{(k+1)}=\left(\frac{n}{\sigma_{(k)}^2}+\frac{1}{S_\alpha}\right)^{-1}\frac1{\sigma_{(k)}^2}\sum_{i=1}^ns_i^{(k)}(y_i-x_i^{\top}\beta_{(k)}),\\
&\beta_{(k+1)}=\left(\frac1{\sigma_{(k)}^2}\sum_{i=1}^ns_i^{(k)}x_ix_i^{\top}+U^{-1}\right)^{-1}\frac1{\sigma_{(k)}^2}\sum_{i=1}^ns_i^{(k)}x_i(y_i-\alpha_{(k+1)}),\\
&\sigma^2_{(k+1)}=\left(2+a+\frac{n}{1+\gamma}\right)^{-1}\left\{a+\sum_{i=1}^ns_i^{(k)}(y_i-\alpha_{(k+1)}-x_i^{\top}\beta_{(k+1)})^2\right\}.
\end{split}
\end{equation*}
\end{itemize} 
We adopt the final values as sampled values from the full conditional distribution.

\item
(Sampling from $u_1,\ldots,u_p$ and $\la$)
\begin{itemize}
\item 
(Laplace prior) \ \  
The full conditional distribution of $1/u_k$ is inverse-Gaussian with parameters $\mu=\sqrt{\la/\beta_k^2}$ and $\delta=\la$ in the parametrization of the inverse-Gaussian density given by
$$
f(x)=\sqrt{\frac{\delta}{2\pi}}x^{-3/2}\exp\left\{-\frac{\delta(x-\mu)^2}{2\mu^2x}\right\}, \ \ x>0.
$$
The full conditional distribution of $\la^2$ is ${\rm Ga}(c_1+p,c_2+\sum_{k=1}^pu_k/2)$.

\item 
(Horseshoe prior) \ \ 
The full conditional distribution of $u_k$, $\xi_k$ and $\la$ are ${\rm IG}(1,\la/\xi_k+\beta_k^2/2)$, ${\rm IG}(1,1+\la/u_k)$ and ${\rm Ga}(c_1+p_2,c_2+\sum_{k=1}^pu_k^{-1}\xi_k^{-1})$, respectively.

\end{itemize}
\end{itemize}

Note that the sampling scheme in the main parameter $\beta$ does not use the previous sampled values of $\beta$ and there is no rejection steps, thereby autocorrelation of $\beta$ generated from the above MCMC algorithm is expected to be very small, which will be demonstrated in our numerical studies in Section \ref{sec:num}.

\section{Numerical studies}\label{sec:num}

\subsection{Bayesian robustness properties}
\begin{figure}[!htb]
\centering
\includegraphics[width=12cm,clip]{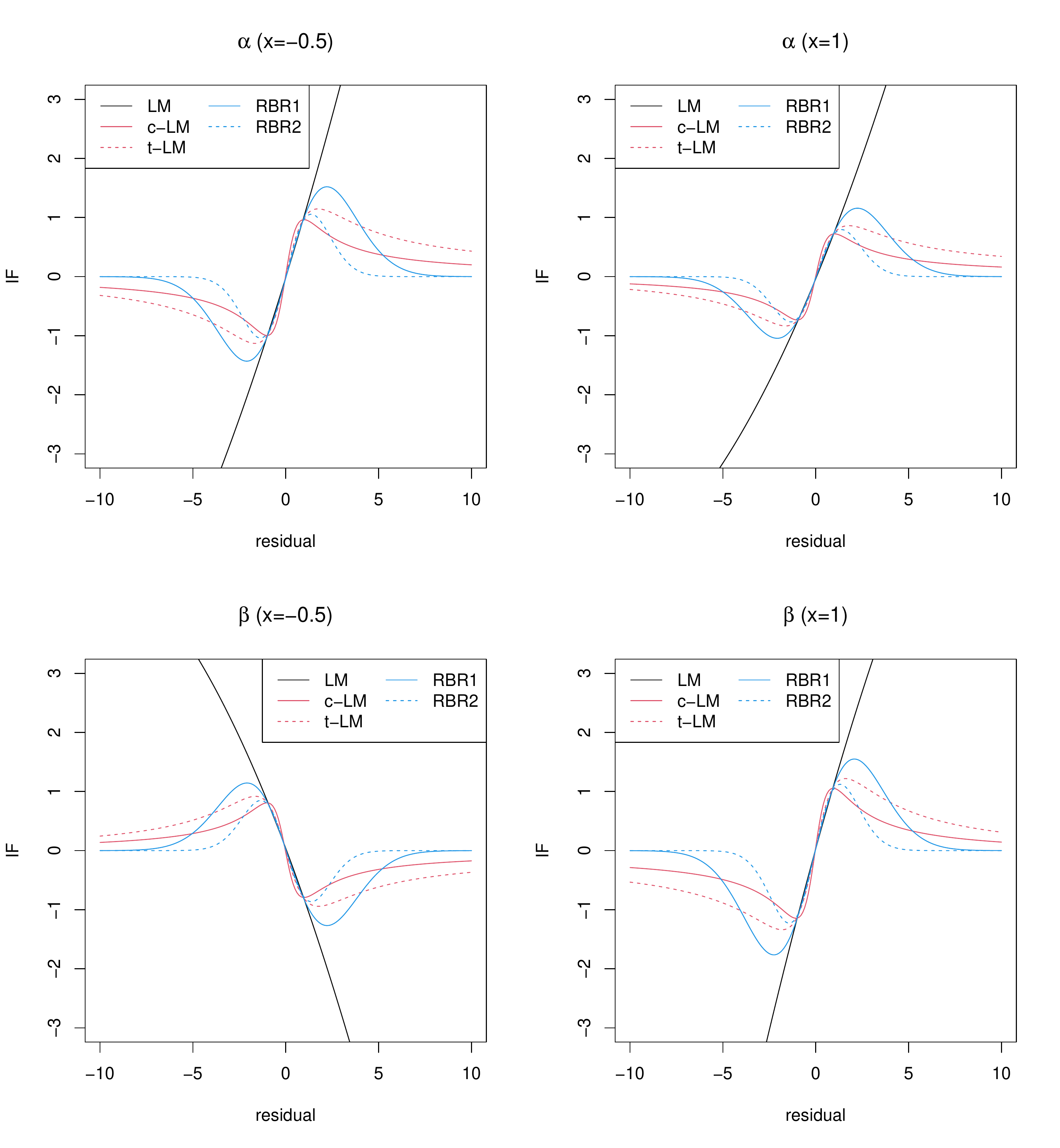}
\caption{Influence functions for $\alpha$ and $\beta$ under a simple linear model.}
\label{fig:IF}
\end{figure}

We demonstrate Bayesian robustness properties of the proposed method compared with existing ones. 
We first consider the influence function of posterior means.
To this end, we employ a simple linear regression model given by 
\begin{equation}\label{slm}
y_i=\alpha+\beta x_i+\ep_i, \ \ \ i=1,\ldots,n,
\end{equation}
where $\alpha=0, \beta=1$, $\ep_i\sim N(0,\sigma^2)$ with $\sigma^2=1$, $x_i$ is generated from the standard normal distribution, and we set $n=300$. 
Let $\theta=(\alpha,\beta,\sigma^2)$ be the set of unknown parameters in the model. 
Let $f(y_i|x_i;\theta)$ be the assumed density function for each $y_i$ given $x_i$. 
Then, the Bayesian analog of the influence function for the posterior means \cite{B1998, Nakagawa2019} of $\theta_k\  (k=1,2,3)$ evaluated at $x$ is given by ${\rm IF}_k(z|x)=n {\rm Cov}_{\theta|D}(\theta_k, H(\theta, z|x))$, where ${\rm Cov}_{\theta|D}$ denotes the covariance with respect to the posterior distribution of $\theta$ under the model (\ref{slm}), and $H(\theta, z|x)$ is the derivative of the (synthetic) likelihood under contamination with respect to the contamination ratio \cite{B1998, Nakagawa2019}.
We used uniform priors for $\alpha$ and $\beta$, and ${\rm Ga}(1,1)$ prior for $\sigma^{-2}$ to obtain the posterior distribution of $\theta$ under the model (\ref{slm}).
Under the standard likelihood function, it holds that $H(\theta, z|x)=\log f(\alpha+\beta x+z|x;\theta)-\int \log f(t|x;\theta) g(t|x)\mathrm{d}t$, where $g(\cdot|x)=\phi(\cdot; x, 1)$ is the true density under (\ref{slm}).
Also, it follows that
$$
H(\theta, z|x)=\frac{1}{\gamma}\left\{\frac{f(\alpha+\beta x+z|x;\theta)^{\gamma}}{\int f(t|x;\theta)^{\gamma} g(t|x;\theta_0)\mathrm{d}t} -1\right\}
$$ 
under the $\gamma$-divergence. 
We note that $z$ can be interpreted as the residual of the outlying value, namely, the distance between the outlying value and the regression line $\alpha+\beta x$.
We approximated the integral appeared in $H$ by Monte Carlo integration based on 2000 random samples from $g(\cdot|x)$.
Based on 10000 posterior samples of $\theta$, we computed ${\rm IF}_1(z|x)$ and ${\rm IF}_2(z|x)$ which are the influence functions for $\alpha$ and $\beta$, respectively, for $x\in \{-0.5, 1\}$ and $z\in [-10, 10]$, under the $\gamma$-divergence with $\gamma=0.2$ (RBR1) and $\gamma=0.5$ (RBR2). 
For comparison, we also computed the influence functions using the normal distribution (LM), Cauchy distribution (c-LM) and $t$-distribution with 3 degrees of freedom (t-LM) for the error term $\ep_i$ in the model (\ref{slm}).
The results are presented in Figure \ref{fig:IF}.
It shows that using heavy-tailed distributions such as Cauchy and $t$-distribution provide bounded influence functions for both parameters, but the influence functions based on the proposed $\gamma$-divergence method quickly converges to $0$ as $|z|$ increases. 
This would indicate more strong robustness of the proposed method than using the heavy-tailed distributions.

We next more directly evaluate the robustness properties.
To this end, we employ the contaminated structure for the error term, $\ep_i\sim N(0,a^2\sigma^2)$ for $i=1,\ldots,n\omega$, and $\ep_i\sim N(0,\sigma^2)$ for $i=n\omega+1,\ldots,n$, so the first $n\omega$ observations are outliers, where $\omega$ and $a$ control the number of outliers and severity of the contamination, respectively.  
We then define the oracle posterior distribution $\pi_{\ast}(\alpha,\beta)$ as the posterior distribution based on the normality assumption $\ep_i\sim N(0,\sigma^2)$ and observations without outliers, that is, $(x_i, y_i)$ for $i=n\omega+1,\ldots,n$. 
Let $\pi(\alpha,\beta)$ be (synthetic) posterior distributions based on the whole data including outliers.
If the distance between $\pi(\alpha,\beta)$ and $\pi_{\ast}(\alpha,\beta)$ is small, we can conclude that the posterior $\pi(\alpha,\beta)$ successfully eliminate the information from outliers to make the posterior inference robust.
Therefore, we assess the distance by computing the Kullback-Leibler divergence given by $\int \pi_{\ast}(\alpha,\beta)\log\{\pi_{\ast}(\alpha,\beta)/\pi(\alpha,\beta)\}\mathrm{d}\alpha \mathrm{d}\beta$.
In Table \ref{tab:KL}, we reported the results averaged over 300 replications under scenarios with $\omega\in \{0.05, 0.1, 0.15, 0.2\}$ and $a\in \{10, 20\}$.
It is observed that the proposed synthetic posterior is reasonably close to the oracle posterior, which is consistent to the theoretical argument given in Section \ref{sec:model}, and the distance is smaller than those of the other methods.

\begin{table}[!htbp]
\caption{Kullback-Leibler divergence averaged over 300 replications.
\label{tab:KL}}
\begin{center}
\begin{tabular}{cccccccccccccc}
\toprule
&& \multicolumn{5}{c}{$a=10$} && \multicolumn{5}{c}{$a=20$}\\
$\omega$ &  & LM & c-LM & t-LM & RBR1 & RBR2 &  & LM & c-LM & t-LM & RBR1 & RBR2 \\
 \midrule
0.05 &  & 1.689 & 0.704 & 0.256 & 0.188 & 0.318 &  & 2.883 & 0.682 & 0.238 & 0.172 & 0.313 \\
0.1 &  & 2.212 & 0.605 & 0.292 & 0.229 & 0.328 &  & 3.471 & 0.552 & 0.257 & 0.171 & 0.297 \\
0.15 &  & 2.886 & 0.587 & 0.422 & 0.341 & 0.386 &  & 4.214 & 0.537 & 0.418 & 0.240 & 0.367 \\
0.2 &  & 3.074 & 0.576 & 0.579 & 0.429 & 0.393 &  & 4.445 & 0.547 & 0.683 & 0.297 & 0.398 \\
\bottomrule
\end{tabular}
\end{center}
\end{table}

\subsection{Simulation study}
We here evaluate the performance of the proposed methods through simulation studies.
We first compare the point and interval estimation performance of the proposed methods with those of some existing methods.
To this end, we consider the following regression model with $n=100$ and $p=20$:
$$
y_i=\alpha+\beta_1x_{i1}+\cdots+\beta_px_{ip}+\ep_i, \ \ \ \ i=1,\ldots,n,
$$
where $\alpha=0.5$, $\beta_1=\beta_4=0.5$ and $\beta_7=\beta_{10}=\beta_{13}=2$ and the other $\beta_k$'s were set to $0$.
The covariates $x_i=(x_{i1},\ldots,x_{ip})$ were generated from a multivariate normal distribution $N_p(0,\Sigma)$ with $\Sigma=(\rho^{|i-j|})_{1\leq i,j\leq p}$ with $\rho=0.2$.
For the error term $\ep_i$, we adopted contaminated structure given by $\ep_i\sim (1-\omega_i)N(0,1)+\omega_i f_c$, where $f_c$ is a contamination distribution and $\omega_i$ is contamination probability which might be heterogeneous over samples.
We considered two settings $f_c$, that is, (I)$f_c\sim N(0,10^2)$ and (II)$f_c\sim N(10,1)$, noting that the contamination distribution has very large variance in scenario (I) while the contamination distribution tends to produce large values in scenario (II). 
Regarding the contamination probability, we considered the following scenarios: 
\begin{align*}
({\rm Homo})& \ \ \omega_i=\omega\in \{0,0.05,0.10,0.15,0.20\}\\
({\rm Hetero})& \ \ \omega_i=\delta\times {\rm logistic}(-3.3+x_{i10}), \ \  \delta\in \{0,1,2,3,4\}.
\end{align*}
Note that the contamination probability in the first setting is constant over the samples, which is refereed to homogeneous contamination.
On the other hand, in the second setting, the probability depends on covariates and is different over the samples, so that it is refereed to heterogenous contamination.

For the simulated dataset, we applied the proposed robust methods with Laplace and horseshoe priors, denoted by RBL and RHS, respectively, as well as the standard (non-robust) Bayesian lasso (BL).
The tuning parameter $\gamma$ in the proposed method is set to $\gamma=0.2$.
Moreover, as existing robust methods, we also applied the regression model with the error term following Cauchy distribution and $t$-distribution with $3$ degrees of freedom, denoted by c-BL and t-BL, respectively. 
In applying the above methods, we generated 2000 posterior samples after discarding the first 1000 samples as burn-in.
For point estimates of $\beta_k$'s, we computed posterior median of each element of $\beta_k$'s, and their performance is evaluated via mean squared error (MSE) defined as $p^{-1}\sum_{k=1}^p(\widehat{\beta}_k-\beta_k)^2$.
We also computed $95\%$ credible intervals of $\beta_k$'s, and calculated average lengths (AL) and coverage probability (CP) defined as $p^{-1}\sum_{k=1}^p|{\rm CI}_k|$ and $p^{-1}\sum_{k=1}^pI(\beta_k\in {\rm CI}_k)$, respectively.
These values were averaged over 300 replications of simulating datasets.

\begin{figure}[!htb]
\centering
\includegraphics[width=12cm,clip]{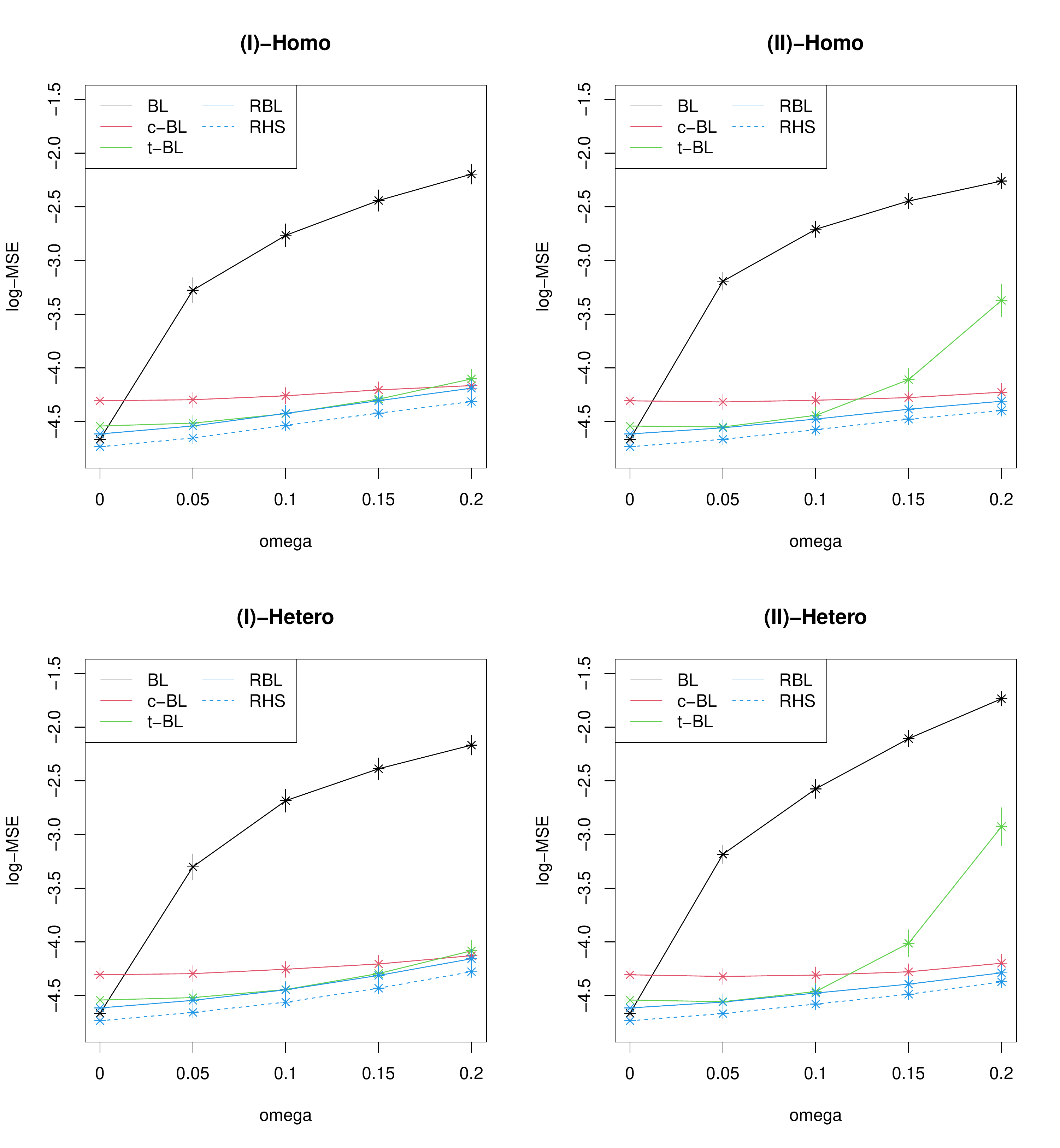}
\caption{Average values of log-MSE with $\pm 3$ Monte Carlo error bands based on 300 replications in four simulation scenarios.}
\label{fig:MSE}
\end{figure}

In Figures \ref{fig:MSE}, we presented the results of logarithm of MSE (log-MSE) with error bars corresponding to three times estimated Monte Carlo errors.
When there is no outliers, the standard BL method performs quite well while the proposed two robust methods (RBL, RHS) are comparable.
On the other hand, the performance of BL gets worse as the ratio of outliers increases, compared with the other robust methods.
Comparing the proposed methods with t-BL, it is observed that they perform similarly when the contaminated error distribution is symmetric and has large variance as in Scenario (I), possibly because $t$-distribution can be effectively adapted to such structure.
However, when the contaminated distribution is not symmetric as in Scenario (II), the performance of t-BL gets worse than the proposed robust methods as the contamination ratio increases.
In these scenarios, c-BL works better than t-BL, but the proposed method still provides better results than c-BL.
It is also observed that c-BL is considerably inefficient compared with the other robust methods when the contamination ratio is not large.
Comparing the proposed two robust methods, RHS is slightly better than RBL in all the scenarios.
As the horseshoe prior is known to have better performance than Laplace prior as shrinkage priors under no contamination, this results would indicate that such property can be inherited even under contamination by using the proposed robust approach.

\begin{figure}[!htb]
\centering
\includegraphics[width=12cm,clip]{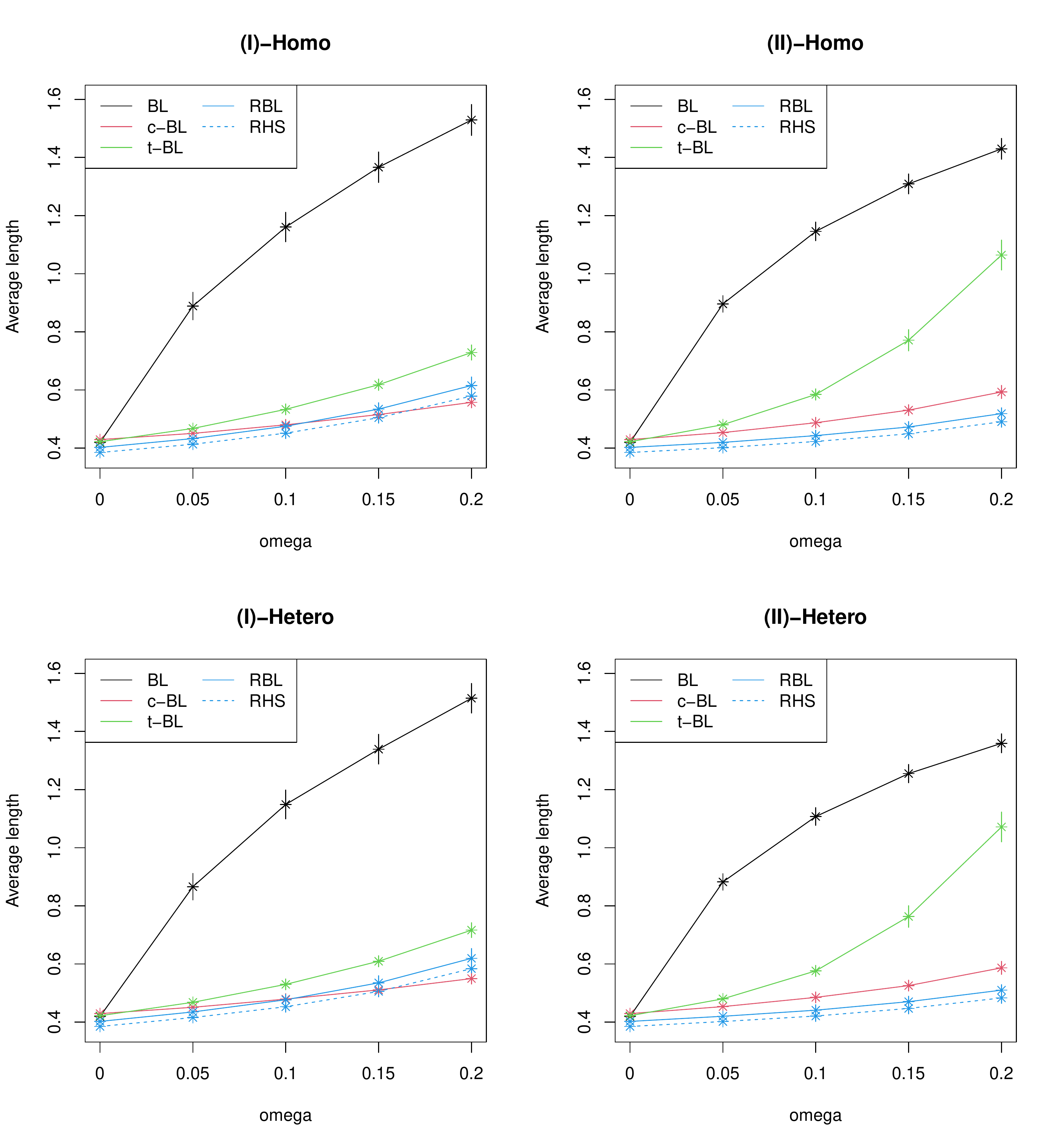}
\caption{Average values of average length of $95\%$ credible intervals with $\pm 3$ Monte Carlo error bands based on 300 replications in four simulation scenarios.}
\label{fig:AL}
\end{figure}

Regarding interval estimation, the results for AL and CP are given in Figure \ref{fig:AL} and Table \ref{tab:cp}, respectively.
From Table \ref{tab:cp}, we can see that CPs of the proposed methods are around the nominal level whereas t-CL and c-BL shows over-coverage and short-coverage properties, respectively, in some scenarios. 
Figure \ref{fig:AL} reveals a similar trend to one observed in log-MSE, so that the credible intervals of BL are shown to be very inefficient when there exist outliers.
Also it is observed that the credible intervals of t-BL and c-BL tend to be inefficient compared with the proposed methods. 
Comparing the proposed two robust methods, RHS provides slightly more efficient interval estimation than RBL, which is consistent with the results of log-MSE in Figure \ref{fig:MSE}.

\begin{figure}[!htb]
\centering
\includegraphics[width=11cm,clip]{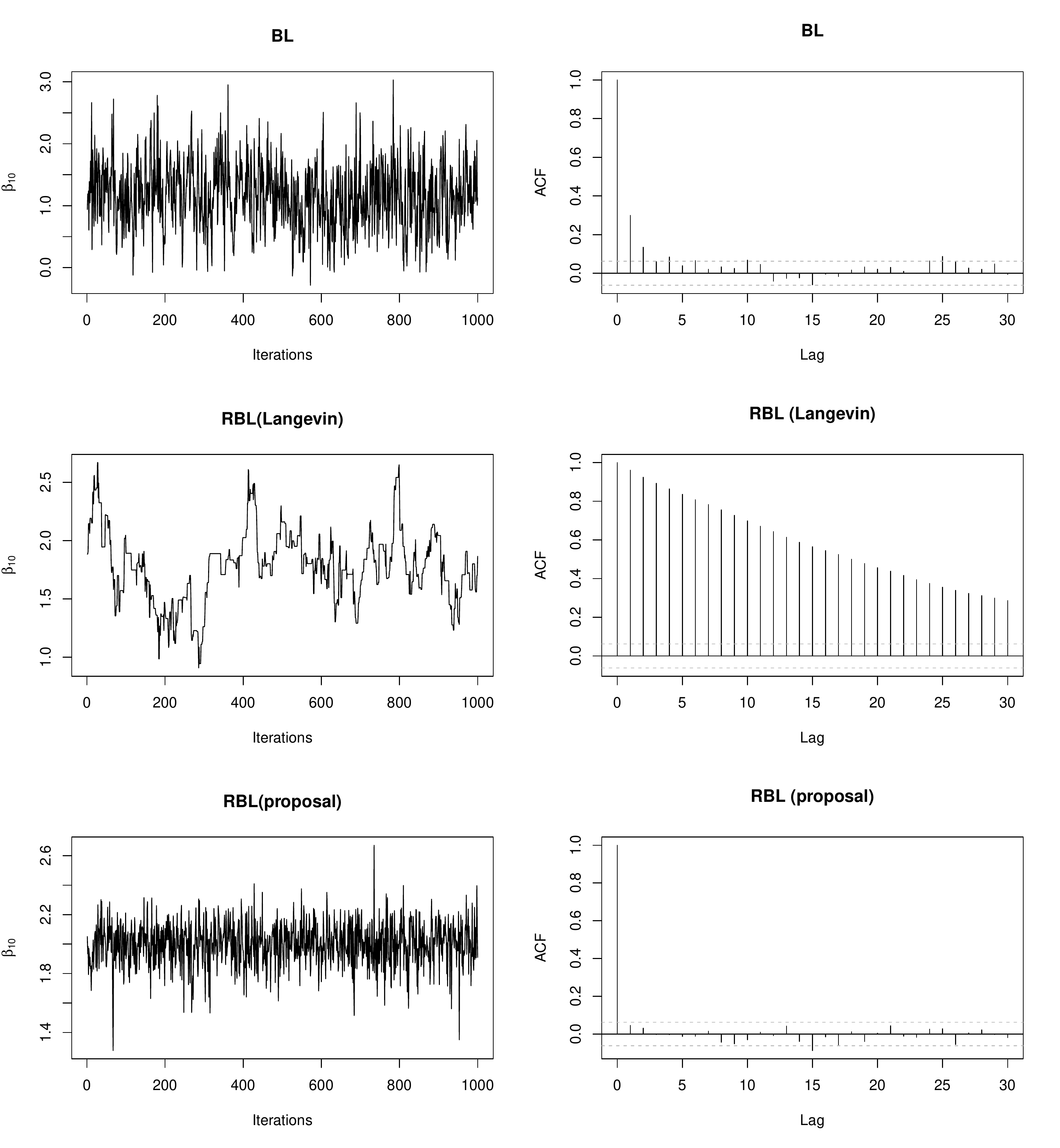}
\caption{Trace plots and autocorrelations of posterior samples of $\beta_{10}$ based on three methods (BL, RBL-Langevin and RBL-proposal) in (II)-Homo case with $\omega=0.20$ and $p=20$}
\label{fig:plot20}
\end{figure}

\begin{figure}[!htb]
\centering
\includegraphics[width=11cm,clip]{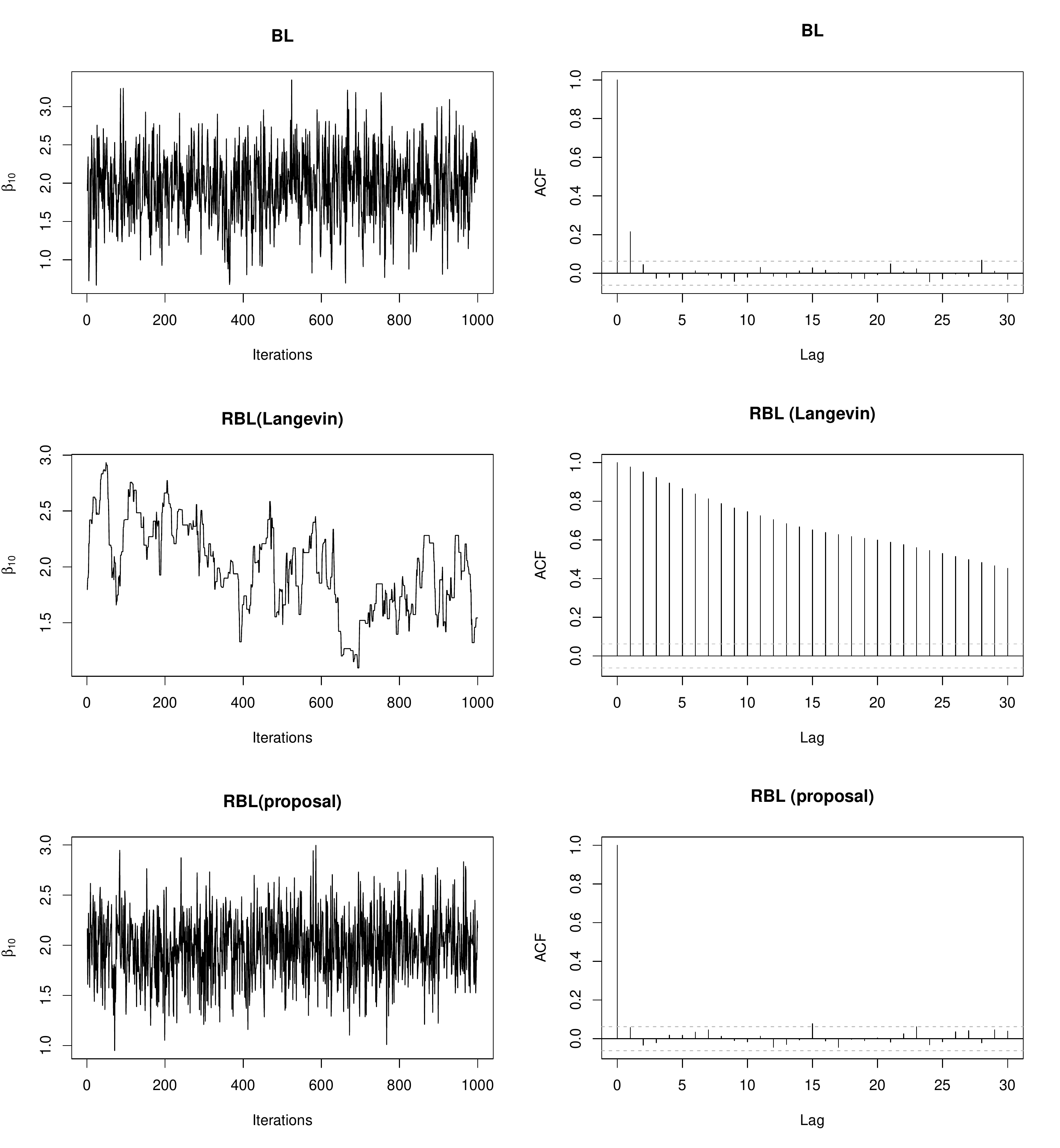}
\caption{Trace plots and autocorrelations of posterior samples of $\beta_{10}$ based on three methods (BL, RBL-Langevin and RBL-proposal) in (II)-Homo case with $\omega=0.20$ and $p=40$}
\label{fig:plot40}
\end{figure}

We next checked mixing properties of the proposed MCMC algorithm (denoted by RBL-proposal) compared with standard methods. We consider the same setting as above simulations. In particular, we show only results in the case of (II)-Homo. Note that we can obtain similar results in other cases, i.e., (I)-Homo, (I)-Hetero and (II)-Hetero. We set $\omega=0.20$. In addition to the case of $p=20$, we also consider the case of $p=40$, where we set the same true non-zero regression coefficients as the case of $p=20$. For comparison, we employed (non-robust) Bayesian lasso (BL) and robust Bayesian lasso with Langevin algorithm \citep[e.g.][]{WT2011} with the step size $0.01$ (denoted by RBL-Langevin). Note that the Langevin algorithm is applied to the full conditional distribution of $\beta$, so that this algorithm has only difference from the proposed algorithm in sampling from the full conditional distribution of $\beta$.
Figure \ref{fig:plot20} shows mixing and autocorrelation results for one-shot posterior simulation of $\beta_{10}$. 
As is well-known, since the ordinal BL have the efficient Gibbs sampling algorithm, we can find that mixing and autocorrelation of them are quite well whereas the sample path of BL is away from the true value due to outliers. 
From the second row Figure \ref{fig:plot20}, it is observed that the Langevin algorithm produces poor mixing and relatively high autocorrelation. 
We tried other choices of tuning parameters in the Langevin algorithm, but the results were comparable. 
On the other hand, the bottom of Figure \ref{fig:plot20} shows that the mixing properties of the proposed algorithm are quite satisfactory, that is, sample paths are around the true value and there is almost no autocorrelations. 
It should be noted that the proposed algorithm does not depend on any tuning parameters unlike the Langevin algorithm algorithm. 
In Figure \ref{fig:plot40}, we reported the results under $p=40$, which also clearly shows the preferable performance of the proposed sampling algorithm compared with the direct application of the Langevin algorithm.

\begin{table}[!htbp]
\caption{Coverage probability of $95\%$ credible intervals averaged over 300 replications.
\label{tab:cp}}
\begin{center}
\begin{tabular}{ccccccccccccccc}
\toprule
&& \multicolumn{5}{c}{(I)-Homo} && \multicolumn{5}{c}{(II)-Homo}\\
$\omega$ &  & BL & c-BL & t-BL & RBL & RHS &  & BL & c-BL & t-BL & RBL & RHS \\
 \midrule
0 &  & 96.0 & 92.6 & 94.8 & 93.5 & 93.4 &  & 96.0 & 92.6 & 94.8 & 93.5 & 93.4 \\
0.05 &  & 96.0 & 93.9 & 96.6 & 94.2 & 94.0 &  & 96.4 & 94.2 & 97.2 & 93.7 & 93.7 \\
0.1 &  & 96.4 & 94.5 & 97.3 & 94.6 & 94.2 &  & 96.5 & 95.4 & 98.5 & 94.0 & 93.9 \\
0.15 &  & 96.6 & 95.4 & 98.1 & 95.2 & 95.2 &  & 96.7 & 96.6 & 99.0 & 94.4 & 94.1 \\
0.2 &  & 96.6 & 96.1 & 98.5 & 96.3 & 96.0 &  & 96.7 & 97.4 & 98.6 & 95.3 & 94.8 \\
\bottomrule
\end{tabular}

\vspace{5mm}
\begin{tabular}{cccccccccccccc}
\toprule
&& \multicolumn{5}{c}{(I)-Hetero} && \multicolumn{5}{c}{(II)-Hetero}\\
$\delta$  &  & BL & c-BL & t-BL & RBL & RHS &  & BL & c-BL & t-BL & RBL & RHS \\
 \midrule
0 &  & 96.0 & 92.6 & 94.8 & 93.5 & 93.4 &  & 96.0 & 92.6 & 94.8 & 93.5 & 93.4 \\
1 &  & 95.9 & 93.4 & 96.3 & 94.3 & 94.1 &  & 95.9 & 94.0 & 97.1 & 93.7 & 93.6 \\
2 &  & 96.0 & 94.8 & 97.6 & 95.1 & 94.9 &  & 95.0 & 95.3 & 98.5 & 94.0 & 93.7 \\
3 &  & 96.7 & 95.1 & 98.4 & 95.5 & 95.3 &  & 94.3 & 96.3 & 98.9 & 94.4 & 94.1 \\
4 &  & 96.6 & 95.6 & 98.8 & 96.2 & 96.2 &  & 92.7 & 96.9 & 96.7 & 95.0 & 94.5 \\
\bottomrule
\end{tabular}
\end{center}
\end{table}

\subsection{Real data examples}
We compare results of the proposed methods with that of the non-robust Bayesian Lasso through applications to two famous datasets, Boston Housing \citep{HR1978} and Diabetes data \citep{Efron2004}.
The response variable in the Boston housing data is corrected median value of owner-occupied homes in USD 1000's, and there are 15 covariates including one binary covariate. 
We standardized 14 continuous valued covariates, and included squared values of these covariates, which results in 29 predictors in our models. 
The sample size is $506$. 
Regarding the Diabetes data, the data contains information of $442$ individuals and we adopted 10 covariates, following \cite{PC2008}.

\begin{figure}[!htb]
\centering
\includegraphics[width=13cm,clip]{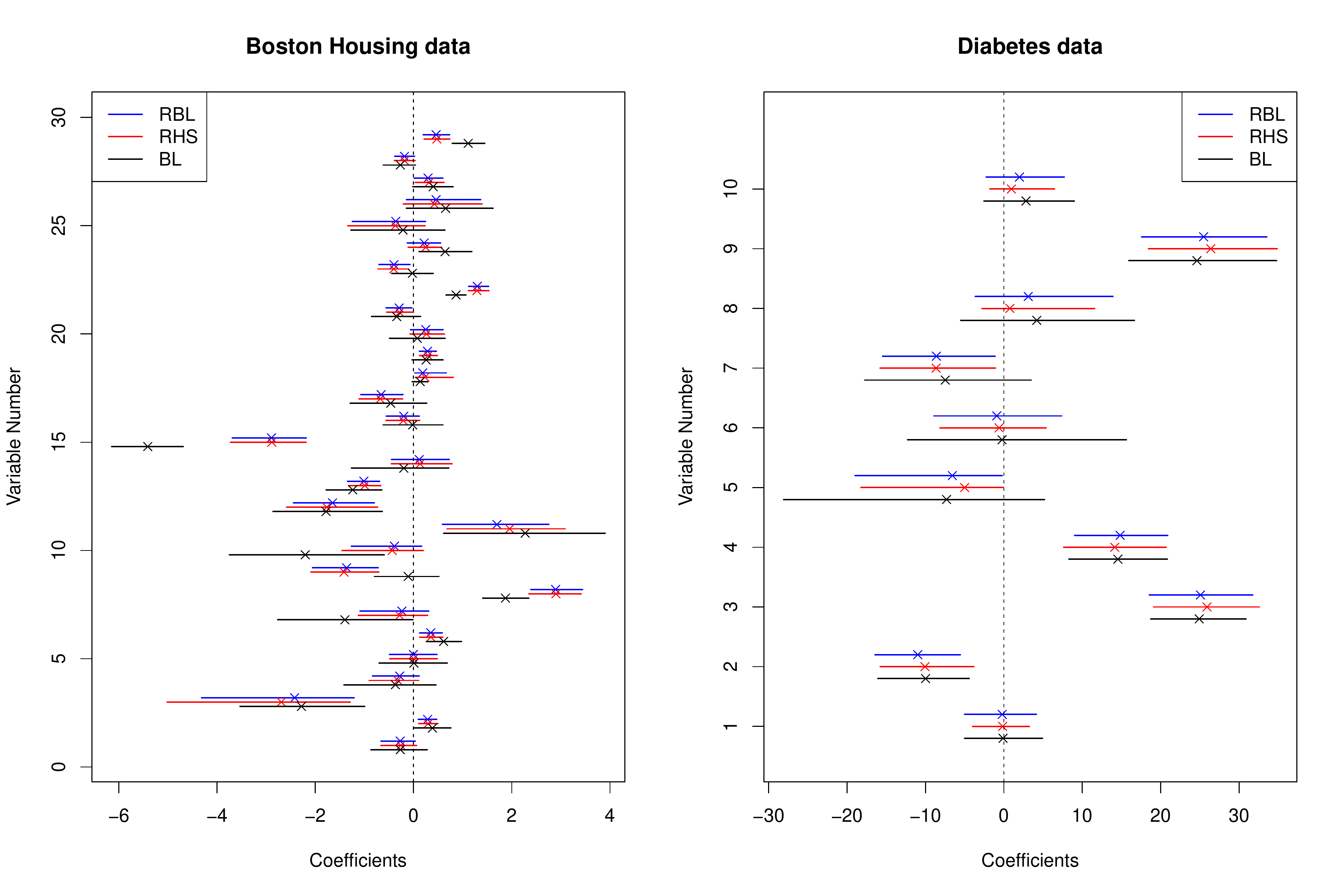}
\caption{$95\%$ credible intervals with posterior medians ($\times$) of regression coefficients based on the standard Bayesian Lasso (BL), the proposed robust Bayesian methods with Laplace prior (RBL) and horseshoe prior (RHS), applied to Boston Housing data and Diabetes data.}
\label{fig:ex}
\end{figure}

For the datasets, we applied the proposed robust Bayesian methods with Laplace prior (RBL) and horseshoe prior (RHS).
We set $\gamma=0.2$ in both methods.
For comparison, we also applied the standard non-robust Bayesian Lasso (BL).
Based on 4000 posterior samples after discarding 1000 posterior samples, we computed posterior medians as well as $95\%$ credible intervals of regression coefficients, which are shown in Figure \ref{fig:ex}.  
It is observed that the two robust methods, RBL and RHS produce similar results while the standard BL provides quite different results than the others in some coefficients.
In particular, in the Diabetes dataset, the two robust methods detected 5th and 7th variables as significant ones based on their credible intervals while the credible intervals of the BL method contains $0$, which shows that the robust methods may be able to detect significant variables that the non-robust method cannot.
A similar phenomena is observed in several covariates in the Boston Housing data.
Comparing two figures, the degree of difference of the results between the two robust methods and the non-robust method in the Diabetes data is smaller than that in the Boston Housing data, for example, the posterior medians among the three methods are almost identical in the Diabetes data.
This might show that the ratio of outliers in the Diabetes dataset is smaller than that of the Boston Housing data.

\section{Conclusions and discussion}\label{sec:dis}
We proposed a new robust Bayesian regression method by using synthetic posterior based on $\gamma$-divergence.
Using a technique of Bayesian bootstrap that optimizes a weighted objective function within Gibbs sampling, we developed an efficient posterior computation algorithm to generate posterior samples of regression coefficients under shrinkage priors.
The numerical performance of the proposed method compared with existing methods are investigated through simulation and real data examples.

Although our presentation is focused on a linear regression in this paper, the proposed method can be conceptionally extended to other models such as generalized linear models. 
However, under generalized linear models, corresponding objective functions from $\gamma$-divergence is not necessarily tractable since it might include intractable integrals or infinite sums \citep{KF2019}, thereby the posterior computation would not be feasible even if we use a similar techniques used in this paper.
Only logistic regression for binary outcomes could be a tractable regression model in which $\gamma$-divergence is obtained in an analytical form, and outliers in logistic regression is typically related to a mislabeling problem \citep[e.g.][]{Hung2018}. 
The detailed investigation including developing an efficient posterior computation algorithm would be an important future work.

Regarding the choice of $\gamma$, it is not straightforward to estimate/select the value in a data-dependent way as it is not a model parameter. 
From the results in Section 3.1, the posterior distribution would not be very sensitive to the different value of $\gamma$, and $\gamma$-divergence is known to be robust as long as $\gamma>0$.
Hence, our recommendation is simply using a small value for $\gamma$ such as $\gamma=0.2$ as adopted in our numerical studies. 
However, it would be quite interesting to consider some data-dependent approaches to the choice of $\gamma$.

\section*{Acknowledgement}
This work is partially supported by Japan Society for Promotion of Science (KAKENHI) grant numbers 18K12757 and 17K14233.

\vspace{1cm}

\bibliographystyle{chicago}
\bibliography{refs}

\end{document}